
\input harvmac

\Title{\vbox{\baselineskip12pt\hbox{CTP/TAMU-49/93}
\hbox{CERN-TH.7088/93}
}}
{\vbox{\centerline {New Black Hole, String and Membrane Solutions}
\vskip2pt\centerline{of the Four-Dimensional Heterotic String*}}}
\footnote{}
{*Work supported in part by NSF grant PHY-9106593.}
\centerline{M.~J.~Duff$^1$, Ramzi R.~ Khuri$^2$\footnote{$^\dagger$}
{Supported by a World Laboratory Fellowship.}, Ruben Minasian$^1$ and Joachim
Rahmfeld$^1$}
\bigskip\centerline{$^1${\it Center for Theoretical Physics}}
\centerline{\it Texas A\&M University}\centerline{\it College Station, TX
77843}
\bigskip\centerline{$^2${\it Theory Division, CERN}}
\centerline{\it CH-1211, Geneva 23, Switzerland}

\vskip .3in
We present solutions of the low-energy four-dimensional heterotic string
corresponding to $p$-branes with $p=0,1,2$, which are characterized by a mass
per unit $p$-volume, ${\cal M}_{p+1}$, and topological ``magnetic'' charge,
$g_{p+1}$. In the extremal limit,  $\sqrt{2} \kappa {\cal M}_{p+1} = g_{p+1}$,
they reduce to the recently discovered non-singular supersymmetric monopole,
string and domain wall solutions. A novel feature is that the solutions involve
both the dilaton and the modulus fields. In particular, the effective scalar
coupling to the Maxwell field, $e^{-\alpha \phi} F_{\mu\nu} F^{\mu\nu}$, gives
rise to a new string black hole with $\alpha = \sqrt{3}$, in contrast to the
pure dilaton black hole solution which has $\alpha=1$. This means that
electric/magnetic duality in $D=4$ may be seen as a consequence of
string/fivebrane duality in $D=10$.

\Date{\vbox{\baselineskip12pt\hbox{CTP/TAMU-49/93}
\hbox{CERN-TH.7088/93}\hbox{November 1993}}}

\lref\chad{J.~M.~Charap and M.~J.~Duff, Phys. Lett. {\bf B69} (1977) 445.}

\lref\dufkhu{M.~J.~Duff and R.~R.~Khuri, CTP/TAMU-17/93 (hep-th/9305142)
       (to appear in Nucl. Phys. B).}

\lref\div{M.~J.~Duff and J.~X.~Lu, CTP/TAMU-54/92 (hep-th@9306052).}

\lref\dkl{M.~J.~Duff, R.~R.~Khuri and J.~X.~Lu, Nucl. Phys. {\bf B377} (1992)
 281.}

\lref\duff{M.~J.~Duff, Class. Quantum Grav. {\bf 5} (1988) 189.}

\lref\fbrane{M.~J.~Duff and J.~X.~Lu, Nucl. Phys. {\bf B354} (1991) 141.}

\lref\dupo{M.~J.~Duff and C.~N.~Pope, Nucl. Phys {\bf B255} (1985) 355.}

\lref\duality{M.~J.~Duff and J.~X.~Lu, Nucl. Phys. {\bf B354} (1991) 129.}

\lref\sgone{E.~Cremmer, S.~Ferrara, L.~Girardello, A.~Van Proeyen,
 Nucl. Phys. {\bf B212} (1983) 413.}

\lref\khuphy{R.~R.~Khuri, Phys. Lett. {\bf B294} (1992) 325.}

\lref\khunu{R.~R.~Khuri, Nucl. Phys. {\bf B387} (1992) 315.}

\lref\khu{R.~R.~Khuri, Phys. Lett. {\bf B259} (1991) 261.}

\lref\gib{G.~W.~Gibbons, Nucl. Phys. {\bf B207} (1982) 337.}

\lref\gibp{G.~W.~Gibbons and M.~J.~Perry, Nucl. Phys. {\bf B248} (1984) 629.}

\lref\gim{G.~W.~Gibbons and K.~Maeda, Nucl. Phys. {\bf B298} (1988) 741.}

\lref\ghs{D.~Garfinkle, G.~T.~Horowitz and A.~Strominger, Phys. Rev. {\bf D43}
(1991) 3140.}

\lref\strom{A.~Strominger, Nucl. Phys. {\bf B343} (1990) 167.}

\lref\host{G.~T.~Horowitz, A.~Strominger, Nucl. Phys. {\bf B360} (1991)
    2930.}

\lref\hor{G.~T.~Horowitz, UCBTH-92-32 (hep-th@9210119).}

\lref\gps{S.~B.~Giddings, J.~Polchinski and A.~Strominger, NSF-ITP-93-62
  (hep-th@9305083).}

\lref\stw{A.~Shapere, S.~Trivedi and F.~Wilczek, Mod. Phys. Lett. {\bf A6}
(1991) 2677.}

\lref\klopv{R.~Kallosh, A.~Linde, T.~Ortin, A.~Peet and A.~Van~Proeyen,
       Phys. Rev. {\bf D46} (1992) 5278.}

\lref\kal{R.~Kallosh, SU-ITP-92-1 (hep-th@9201029).}

\lref\ko{R.~Kallosh and T.~Ortin, SU-ITP-93-3 (hep-th@9302109).}

\lref\dom{P.~Dobiasch and D.~Maison, Gen. Rel. Grav. {\bf 14} (1982) 231.}

\lref\chod{A.~Chodos and S.~Detweiler, Gen. Rel. Grav. {\bf 14} (1982) 879.}

\lref\pol{D.~Pollard, J. Phys. {\bf A16} (1983) 565.}

\lref\grosp{D.~J.~Gross and M.~J.~Perry, Nucl. Phys. {\bf B226} (1983) 29.}

\lref\sork{R.~D.~Sorkin, Phys. Rev. Lett. {51} (1983) 87.}

\lref\sennu{A.~Sen, Nucl. Phys. {\bf B404} (1993) 109.}

\lref\sen{A.~Sen, TIFR-TH-92-57 (hep-th@9210050).}

\lref\sentwo{A.~Sen, TIFR-TH-93-03 (hep-th@9302038).}

\lref\senph{A.~Sen, Phys. Lett. {\bf B303} (1993) 22.}

\lref\schw{J.~H.~Schwarz and A.~Sen, NSF-ITP-93-46,
            CALT-68-1863, TIFR-TH-93-19 (hep-th@9304154).}

\lref\bin{P.~Binetruy, Phys. Lett {\bf 315} (1993) 80.}

\lref\hw{C.~F.~E.~Holzhey and F.~Wilczek, Nucl. Phys. {\bf B360} (1992) 447.}

\lref\monol{C.~Montonen and D.~Olive, Phys. Lett. {\bf B72} (1977) 117.}

\lref\ghl{J.~P.~Gauntlett, J.~A.~Harvey and J.~T.~Liu, EFI-92-67, IFP-434-UNC
   (hep-th@9211056).}

\lref\dab{A.~Dabholkar, G.~Gibbons, J.~A.~Harvey, F.~R.~Ruiz, Nucl. Phys.
        {\bf B340} (1990) 33}

\lref\schwarz{J.~H.~Schwarz, CALT-68-1879 (hep-th@9307121).}

\lref\dauria{R.~D'Auria, S.~Ferrara and M.~Villasante, CERN-TH 6914/93,
      POLFIS-TH 04/93, UCLA/93/TEP/18 (hep-th@9306125).}

\lref\schen{J.~H.~Schwarz and A.~Sen, Phys. Lett. {\bf B312} (1993) 105}

\def\sqr#1#2{{\vbox{\hrule height.#2pt\hbox{\vrule width
.#2pt height#1pt \kern#1pt\vrule width.#2pt}\hrule height.#2pt}}}

\def\a {\alpha}

\newsec{Introduction}
In recent work \dufkhu, supersymmetric soliton solutions of the
four-dimensional heterotic string were presented, describing monopoles,
strings and domain walls. These solutions admit the $D=10$ interpretation of
a fivebrane wrapped around 5, 4 or 3 of the 6 compactified
dimensions and are arguably exact to all orders in $\alpha'$. In this paper,
we extend all three solutions to two-parameter solutions of the low-energy
equations of the four-dimensional heterotic string, characterized by a mass
per unit $p$-volume, ${\cal M}_{p+1}$,  and topological ``magnetic'' charge,
$g_{p+1}$, where $p = 0, 1$ or 2. We show them to be special cases of the
generic black $p$-branes discussed in \div.  The ``neutral'' (as opposed to
the ``gauge'' or ``symmetric'') solitons discussed in \dufkhu\ are recovered
in the extremal limit, $ \sqrt{2} \kappa {\cal M}_{p+1} =  g_{p+1}$, and are
non-singular in the sense that the curvature
singularity disappears when
expressed in terms of the dual $\sigma$-model metric \refs{\div,\dkl}. The
two-parameter solution extending the supersymmetric monopole corresponds to
a magnetically charged black hole, while the solution extending the
supersymmetric domain wall corresponds to a black membrane. By contrast, the
two-parameter string solution does not possess a finite horizon and
corresponds to a naked singularity.

All three solutions involve both the
dilaton and the modulus fields, and are thus to be contrasted with pure dilaton
solutions. In particular, the effective scalar coupling to the Maxwell field,
$e^{-\alpha\phi} F_{\mu\nu} F^{\mu\nu}$, gives rise to a new string black hole
with $\alpha = \sqrt{3}$, in contrast to the pure dilaton solution of the
heterotic string which has $\alpha = 1$
\refs{\gib\gim\ghs\stw\klopv\kal\ko\hor\sen{--}\gps}.
It thus resembles the black hole
previously studied in the context of Kaluza-Klein theories
\refs{\dom\chod\pol\grosp\sork{--}\gibp,\gib,\hor} which also
has $\alpha = \sqrt{3}$, and which reduces to the Pollard-Gross-Perry-Sorkin
\refs{\pol\grosp{--} \sork} magnetic monopole in the extremal limit.
In this connection, we recall the
recent paper of Holzhey and Wilczek \hw, according to which
$\alpha > 1$ black holes behave like elementary particles!

The fact that the heterotic string admits $\alpha = \sqrt{3}$ black holes also
has implications for string/fivebrane duality \refs{\duff\strom{--}\duality}.
We shall show that electric/magnetic duality in $D=4$ may be seen as a
consequence of string/fivebrane
duality in $D=10$.

\newsec{The Solutions}
We begin with the two-parameter black hole.  Inspired by the wrapping
of a fivebrane around five  of the six compactified dimensions
$(x_5, x_6, x_7, x_8, x_9)$, it was shown in \dufkhu\ that the tree-level
effective action for the $D=10$ heterotic string may be reduced to the
following four-dimensional form
\eqn\monact{S_{1}={1\over 2\kappa^2}\int d^4 x \sqrt{-g} e^{-2\Phi -
\sigma_{1}}
\left( R + 4(\partial\Phi)^2 + 4\partial\sigma_{1}\cdot\partial\Phi -
{1\over 4} e^{2\sigma_{1}} F_{\mu\nu} F^{\mu\nu}\right),}
where $\mu,\nu=0,1,2,3$. Here $g_{\mu\nu}$ is the string
sigma-model metric and $\Phi$ is the dilaton. In the case of toroidal
compactification, with $N=4$ supersymmetry in $D=4$, $\sigma_{1}$ is a modulus
field, $g_{44}=e^{-2 \sigma_1}$, and
$F_{\mu\nu}=H_{\mu\nu4}$ where $H=dB$ and $B$ is the
string antisymmetric tensor. However, actions of this type also appear in a
large class of $N=1$ supergravity theories \sgone.
The solution is given by
\eqn\bmonrs{\eqalign{e^{-2\Phi}&=e^{2\sigma_{1}}=\left(1 - {r_-\over r}
\right),\cr
ds^2&=-\left(1-{r_+\over r}\right)\left(1-{r_-\over r}\right)^{-1}dt^2 +
\left(1-{r_+\over r}\right)^{-1}dr^2 +
r^2\left(1-{r_-\over r}\right)d\Omega_2^2,\cr
F_{\theta\varphi}&=\sqrt{r_+r_-} \sin\theta \cr}}
where here, and throughout this paper, we set the dilaton vev $\Phi_0$ equal to
zero.
This represents a magnetically charged black hole with
event horizon at $r=r_+$ and inner horizon at $r=r_-$.
The magnetic charge and mass of the black hole are given by
\eqn\massch{\eqalign{g_{1}&={4\pi\over {\sqrt{2}\kappa}}(r_+r_-)^{{1}\over{2}},
   \cr
{\cal M}_1&={{2 \pi}\over {{\kappa}^2}}(2r_+ - r_-) \cr}}
Changing coordinates via $y=r-r_-$ and taking the extremal limit $r_+=r_-$
yields:
\eqn\monex{\eqalign{e^{2\Phi}&=e^{-2\sigma_{1}}=\left(1 + {r_-\over y}
\right),\cr
ds^2&=-dt^2 + e^{2\Phi}\left(dy^2 + y^2d\Omega_2^2\right),\cr
F_{\theta\varphi}&=r_- \sin\theta.\cr}}
which is just the neutral (i.e. no Yang-Mills) version of the supersymmetric
monopole solution \refs{\khu\khuphy\khunu{--}\ghl,\dufkhu} which saturates
the Bogomol'nyi bound $\sqrt{2} \kappa{\cal M}_1\ge g_1$.

Next we derive a two-parameter string solution which, however,
does not possess a finite event horizon and consequently cannot be
interpreted as a black string. This is inspired by the wrapping of the
fivebrane around four of the compactified dimensions $(x_6, x_7, x_8,x_9)$.
The action is given by
\eqn\stact{S_{2}={1\over 2\kappa^2}\int d^4 x \sqrt{-g} e^{-2\Phi -
2\sigma_{2}}
\left( R + 4(\partial\Phi)^2 + 8\partial\sigma_{2}\cdot\partial\Phi +
2(\partial\sigma_{2})^2 - {1\over 2} e^{4\sigma_{2}} F_\mu F^\mu \right),}
In the case of the torus, $\sigma_2$ is the modulus field
$g_{44}=g_{55}=e^{-2\sigma_2}$ and
$F_{\mu}=H_{\mu45}$. A two-parameter family of solutions is now
given by
\eqn\stsol{\eqalign{e^{2\Phi}&=e^{-2\sigma_{2}}=
(1+k/2-\lambda \ln y),\cr
ds^2&=-(1+k)dt^2 + (1+k)^{-1}(1+k/2-\lambda \ln y)dy^2 +
y^2(1+k/2-\lambda \ln y)d\theta^2 + dx_3^2,\cr
F_\theta&=\lambda \sqrt{1+k},\cr}}
where for $k=0$ we recover the supersymmetric string soliton solution of
\dufkhu\ which is dual to the elementary string solution of Dabholkar {\it et
al}
\dab. The solution shown in \stsol\ in fact represents a naked singularity,
since the event horizon is pushed out to $r_+=\infty$, which agrees with
the Horowitz-Strominger ``no-$4D$-black-string'' theorem \host.

Finally, we consider the two-parameter black membrane solution. In this case,
we wrap the fivebrane around three of the compactified dimensions $(x_7, x_8,
x_9)$.
However, the four-dimensional action necessary to yield membrane solutions
is not obtained by a simple dimensional reduction of the
ten-dimensional action because of the non-vanishing of $F=H_{456}$.
Instead, the effective action is obtained by treating $F^2$ as a cosmological
constant and is  given by
\eqn\memact{S_{3}={1\over 2\kappa^2}\int d^4 x \sqrt{-g}
e^{-2\Phi - 3\sigma_{3}}
\left( R + 4(\partial\Phi)^2 + 12\partial\sigma_{3}\cdot\partial\Phi +
6(\partial\sigma_{3})^2 - e^{6\sigma_{3}} {1\over 2} F^2 \right),}
In the case of the torus, $\sigma_3$ is the modulus field
$g_{44}=g_{55}=g_{66}=e^{-2\sigma_3}$.
The two-parameter black membrane solution is then
\eqn\bmemrs{\eqalign{e^{-2\Phi}&=e^{2\sigma_{3}}=
\left(1-{r\over r_-}
\right),\cr
ds^2&=-\left(1-{r\over r_+}\right)\left(1-{r\over r_-}\right)^{-1}dt^2 +
\left(1-{r\over r_+}\right)^{-1}\left(1-{r\over r_-}\right)^{-4}dr^2 +
dx_2^2 + dx_3^2,\cr
F&=-(r_+r_-)^{-1/2}.\cr}}
This solution represents a black membrane with event horizon at
$r=r_+$ and inner horizon at $r=r_-$.
Changing coordinates via $y^{-1}=r^{-1}-r_{-}^{-1}$ and taking the extremal
limit yields
\eqn\domex{\eqalign{e^{2\Phi}&=e^{-2\sigma_3}=\left(1+{y\over r_-}
\right),\cr
ds^2&=-dt^2 + dx_2^2 + dx_3^2 + e^{2\Phi} dy^2,\cr
F&=-{1\over r_-}.\cr}}
which is just the supersymmetric domain wall solution \dufkhu.

\newsec{Consistency Check}

To verify that the above field configurations are indeed solutions,
we recall the generic
$D$-dimensional black $p$-branes of \div. Consider an antisymmetric tensor
potential of rank $d$, $A_{\mu_1, \ldots, \mu_d}$, in $D$ spacetime dimensions
($\mu = 0, 1, \ldots, (D-1)$) interacting with the canonical Einstein metric
$g_{\mu \nu}(\rm can)$ and a scalar field $\phi$ via the action
\eqn\genact{I_{D}(d)={1\over 2\kappa^2}\int d^D x \sqrt{-g}
\left( R - {1 \over 2}(\partial\phi)^2 -
{1\over 2(d+1)!} e^{-\alpha \phi} F_{\mu_1\mu_2\ldots\mu_{d+1}}
F^{\mu_1\mu_2\ldots\mu_{d+1}}\right)}
where
\eqn\coupl{\alpha^{2}(d) = 4 - {2 d \tilde d \over{d + \tilde d}}}
and where we have introduced the dual worldvolume dimension $\tilde d$ via
\eqn\tildad{\tilde d = D - d - 2.}
To understand this choice of $\alpha$, we recall the action for an
elementary $d$-dimensional extended object (a ``$(d - 1)$-brane'') with
worldvolume coordinates $\xi^{i}  (i=1,2,\ldots,d-1)$, metric
$\gamma_{ij}(\xi)$ and tension $T_{d}$, coupled to the $D$-dimensional
background fields $g_{\mu\nu}({\rm can})$,
$A_{\mu_{1}\cdots\mu_{d}}$ and $\phi$:
\eqn\sigg{\eqalign {S_{d}={T_{d}}\int d^d \xi
&\left( -{1 \over 2}\sqrt{-\gamma} \gamma^{ij}\partial_{i} {\rm X}^{\mu}
\partial_{j} {\rm X}^{\nu} g_{\mu\nu}  e^{{\alpha\phi}/d} +
{(d-2) \over 2}\sqrt{-\gamma} \right. \cr
&\left.-{1\over d!}\epsilon^{i_1 i_2 \cdots i_d} \partial_{i_1} {\rm X}^{\mu_1}
\partial_{i_2} {\rm X}^{\mu_2} \cdots \partial_{i_d} {\rm X}^{\mu_d}
A_{\mu_{1}\cdots\mu_{d}} \right). \cr }}
The above action is the bosonic sector of a spacetime supersymmetric,
$\kappa$-symmetric, Green-Schwarz action for the super $p$-brane.
In particular, the
crucial relative coefficient of the Wess-Zumino and  kinetic
terms (the ``charge
to mass ratio'') is fixed by $\kappa$-symmetry.
The physical significance of the choice of $\alpha$ given in \coupl\ is that
it is the one singled out by the requirement that the combined system
$I_{D}(d)+S_d$ admits elementary $(d-1)$-brane solutions \div\ with mass per
unit $p$-volume, ${\cal M}_{d}$, and Noether ``electric'' charge, $e_{d}$,
given
by $\sqrt{2} \kappa {\cal M}_{d} = e_{d}$. These extreme solutions preserve
half
the spacetime supersymmetries since this charge-to-mass ratio is the one
singled out by $\kappa$-symmetry.

Having fixed the choice of $\alpha$, we now look for black
$(\tilde{d}-1)$-brane solutions of $I_{D}(d)$ alone, carrying a topological
``magnetic'' charge $g_{\tilde{d}}$ obeying the Dirac quantization rule
\eqn\dirac{e_{d}g_{\tilde{d}}=2\pi n,\,\,\,\,\,\,\,\,\,\,\,     n=\rm integer}
The solutions are \div\ (with $i=1,...,\tilde{d}-1$):
\eqn\bpbrs{\eqalign{e^{-2\phi}&= \Delta_{-}^{\alpha},\cr
ds^2&=-\Delta_+ \Delta_-^{-\tilde d \over (d + \tilde d)} dt^2 +
\Delta_+^{-1} \Delta_-^{{\alpha^2 \over 2d} - 1}  dr^2 +
r^2 \Delta_-^{\alpha^2 \over {2d}} d\Omega_{d+1}^2 +
\Delta_-^{d \over (d + \tilde d)} d x^i d x_i,\cr
F_{d+1}&=d(r_+ r_-)^{d/2} \epsilon_{d+1} .}}
 where $\Delta_{\pm} = \left[1 - ({r_{\pm}\over r})^d\right]$
and $\epsilon_{n}$ the volume form on $S^n$.
The magnetic charge $g_{\tilde d}$ and the mass per unit volume
${\cal M}_{\tilde d}$
are related to $r_{\pm}$ by
\eqn\massch{\eqalign{g_{\tilde d}&={\Omega_{d+1} \over{\sqrt{2}\kappa}}
d (r_- r_+)^{d \over2},\cr
{\cal M}_{\tilde d}&={\Omega_{d+1} \over{2 \kappa^2}}
\left((d+1)r_+^d - r_-^d\right)\cr}}
where $\Omega^n$ is the volume of $S^n$.
In the extremal limit, $r_-=r_+$, the solutions saturate the Bogomol'nyi bound
$\sqrt{2} \kappa {\cal M}_{\tilde d} =  g_{\tilde d}$, and also preserve half
the
spacetime supersymmetry. In this limit, the line element reduces to
(with $\beta=0,...,\tilde{d}-1$)
\eqn\lele{ds^2=\Delta_-^{d \over (d + \tilde d)} d x^\beta d x_\beta +
\Delta_-^{\alpha^2 \over 2d}(\Delta_-^{-2} dr^2 + r^2 d\Omega_{d+1}^2 )}
and, for $\alpha \not= 0$, exhibits a curvature singularity at $r=r_-$.
However, as discussed in \refs{\div,\dkl}, when expressed in terms of the
{\it dual}
$(d-1)$-brane $\sigma$-model metric $g_{\mu\nu}(\sigma{\rm-model})=
e^{{\alpha\phi}/d} g_{\mu\nu}({\rm can})$, we find
\eqn\leldd{ds^2=\Delta_-^{(d-2) \over d} d x^\beta d x_\beta +
(\Delta_-^{-2} dr^2 + r^2 d\Omega_{d+1}^2 ) }
which is regular at $r=r_-$. So these extremal black $p$-branes are
non-singular when viewed as solutions of the dual theory. They are also
non-singular in the sense that the proper time taken for a test $p$-brane
to fall radially into a dual source $\tilde{p}$-brane is infinite, and vice
versa \dkl\ (assuming $p, \tilde{p} \geq 0$).

The $\sigma$-model metric is, by definition,  the one for which $S_d$ is
independent of $\phi$. Regarding $S_d$ as a ``matter Lagrangian''
therefore defines a Brans-Dicke type theory with parameter \div\
\eqn\omm{\omega = -{{(D-1)(d-2)-d^2} \over {(D-2)(d-2)-d^2}} }

We now wish to demonstrate that the black hole, string and domain wall
solutions of the heterotic string discussed in section 1 are nothing but
the $({\tilde d}=1, d=1, \alpha=\sqrt{3}, \omega=-{4/3})$,
$({\tilde d}=2, d=0, \alpha=2, \omega=-{3/ 2})$
and $({\tilde d}=3, d=-1, \alpha=\sqrt{7}, \omega=-{10/ 7})$
special cases of the above solutions\footnote{$^\dagger$}
{This means, in particular, that the extreme magnetic
monopole will take an infinite proper time to fall radially into the extreme
electric monopole, and vice versa.}.
To confirm this, it is sufficient to
show that the three actions $S_1$, $S_2$ and $S_3$ may be cast
into the form \genact. This is achieved by transforming the metric
$g_{\mu \nu}$ and scalars $\Phi$ and $\sigma_{i}, (i=1, 2, 3),$ to the
canonical metric
$g_{\mu \nu}(\rm can)$ and scalars $\phi$ and $\lambda$ via the following field
redefinitions:

{\it monopole}
\eqn\one{\eqalign{g_{\mu \nu}&=
e^{{1 \over {\sqrt{3}}}({\sqrt{2}}{\lambda}-\phi)}g_{\mu \nu}({\rm can}),
\cr
\Phi&={1 \over {2 \sqrt{3}} } \left({\lambda \over \sqrt{2}}-2\phi \right), \cr
\sigma_{1}&={1 \over \sqrt{3} } \left({\lambda \over \sqrt{2}}+\phi\right) }}

{\it string}
\eqn\two{\eqalign{g_{\mu \nu}&=
e^{\lambda}g_{\mu \nu}({\rm can}), \cr
\Phi&={1\over 2} (\lambda-\phi), \cr
\sigma_{2}&={1 \over 2}\phi }}

{\it membrane}
\eqn\three{\eqalign{g_{\mu \nu}&=
e^{{1\over \sqrt{7}}({\sqrt{6}}{\lambda}+\phi)}g_{\mu \nu}({\rm can}),
 \cr
\Phi&=
{1\over {2\sqrt{7}}} \left({3\sqrt{3}\over \sqrt{2}}{\lambda}-2\phi \right),
\cr
\sigma_{3}&={1\over \sqrt{7}} \left(-{\lambda\over \sqrt{6}}+\phi \right) }}
Having done this, we can then set $\lambda=0$ to obtain the desired result.
Note that by analytically continuing the solution \bpbrs\ to the cases $d=0$
and $d=-1$, we are extrapolating the meaning of the ADM mass and topological
charge to non-asymptotically flat spacetimes.

\newsec{$\a=\sqrt{3}$ in the Type II String: Electric/Magnetic Duality
    in $D=4$ from Particle/Sixbrane Duality in $D=10$}

We note that the black hole solution corresponds to a
Maxwell-scalar coupling $e^{-\a \phi} F_{\mu \nu} F^{\mu \nu}$ with
$\a=\sqrt{3}$. This is to be contrasted with
the pure dilaton black hole solutions of the heterotic string
that have attracted much attention
recently \refs{\gib\gim\ghs\stw\klopv\kal\ko\hor\sen{--}\gps}
and have $\alpha=1$
\footnote{$^\dagger$}{Contrary to some claims in the
literature, the pure Reissner-Nordstr\"om
black hole with $\a=0$ is also a solution of the low energy heterotic string
equations. This may be seen by noting that it provides a solution to $(N=2,
\, D=4)$ supergravity which is a consistent truncation of toroidally
compactified $N=1, \, D=10$ supergravity \dupo.}. The case $\alpha=\sqrt{3}$
also occurs when
the Maxwell field and the scalar field $\phi$ arise
from a Kaluza-Klein reduction of
pure gravity from $D=5$ to $D=4$:
\eqn\redmet {\hat{g}_{MN} = e^{\phi\over{\sqrt{3}}}
   \pmatrix { g_{\mu\nu}+e^{-\sqrt{3} \phi} A_{\mu}A_{\nu} &
                    e^{-\sqrt{3} \phi} A_{\mu} \cr
          e^{-\sqrt{3} \phi} A_{\nu} &   e^{-\sqrt{3} \phi} \cr}}
where $\hat{g}_{MN} \, (M,N=0,1,2,3,4)$ and $g_{\mu\nu} \,
(\mu,\nu=0,1,2,3)$ are the canonical metrics in 5 and 4 dimensions
respectively. The resulting action is given by
\eqn\redact{S={1\over2 \kappa^2} \int d^4 x \sqrt{-g} \left[ R -{1\over 2}
   (\partial \phi)^2 -{1\over 4} e^{-\sqrt{3} \phi} F_{\mu \nu} F^{\mu \nu}
   \right]}
and it admits as an ``elementary'' solution the $\a=\sqrt{3}$ black hole
metric \bmonrs,  but with the scalar field
\eqn\die{e^{-2 \phi}= \Delta_{-}^{\sqrt{3}}}
and the electric field
\eqn\fse{{1\over \sqrt{2} \kappa}e^{-\sqrt{3} \phi \, \ast} F_{\theta\varphi}=
         {e\over 4 \pi} \sin\theta}
corresponding to an electric monopole with Noether charge $e$.
This system also admits the topological magnetic solution with
\eqn\dim{e^{-2 \phi}= \Delta_{-}^{-\sqrt{3}}}
and the magnetic field
\eqn\fsm{{{1}\over{\sqrt{2}\kappa}} F_{\theta\varphi}={g\over4 \pi}\sin\theta}
corresponding to a magnetic monopole with  topological magnetic
charge $g$ obeying the Dirac quantization rule
\eqn\diqua{eg=2 \pi n, \, \, \, n=\rm{integer}}
In effect, it was for this reason that the $\a=\sqrt{3}$ black hole
was identified as a solution of the Type II string in \div, the fields
$A_{\mu}$ and $\phi$ being just the abelian gauge field and the dilaton of
($N=2,\, D=10$) supergravity which arises from Kaluza-Klein compactification
of ($N=1,\, D=11$) supergravity.

Some time ago, Gibbons and Perry \gibp\ pointed out that $N=8$ supergravity,
compactified from $D=5$ to $D=4$, admits an infinite tower of elementary states
with mass $m_n$ and electric charge $e_n$ given by $\sqrt{2} \kappa m_n=e_n$,
where $e_n$ are quantized in terms of a fundamental charge $e$, $e_n=n \, e$,
and that these elementary states fall into $N=8$ supermultiplets. They also
pointed out that this theory admits an infinite tower of solitonic states
with the masses  $\tilde{m}_n$ and magnetic  charge $g_n$ given by
$\sqrt{2}\kappa\tilde{m}_n=g_n=n\, g$, where $e$ and $g$ obey $eg=2 \pi$,
which also fall into the same $N=8$ supermultiplets. They conjectured,
$\acute{\rm a}$ la Olive-Montonen \monol, that there should
exist a dual formulation of the theory for which the roles of electric
elementary states and magnetic solitonic states are interchanged.
It was argued in \div \ that this electric/magnetic duality conjecture
in $D=4$ could be reinterpreted as a particle/sixbrane duality conjecture
in $D=10$.

To see this, consider the action dual to $S$, with $\a=-\sqrt{3}$, for
which the roles of Maxwell field equations and Bianchi identities are
interchanged:
\eqn\duact{\tilde{S}=
        {1\over2 \kappa^2} \int d^4 x \sqrt{-g} \left[ R -{1\over 2}
   (\partial \phi)^2 -{1\over 4} e^{\sqrt{3} \phi}
      \tilde{F}_{\mu \nu} \tilde{F}^{\mu \nu}
   \right],}
where
$$ \tilde{F}_{\mu\nu} =e^{-\sqrt{3} \phi \, \ast} F_{\mu\nu} $$
This is precisely the action obtained by double dimensional reduction of
a dual formulation of ($D=10, \, N=2$) supergravity in which the two-form
$F_{MN}$ ($M,N=0,...,9$) is swapped for an 8-form
$\tilde{F}_{M_1..M_8}$, where $\tilde{F}_{\mu\nu}=\tilde{F}_{\mu\nu456789}$.
This dual action also admits both electric and magnetic monopole solutions
but because the roles of field equations and Bianchi identities are
interchanged, so are the roles of electric and magnetic. Since the 1-form
and 7-form potentials, which give rise to these 2-form and 8-form
field strengths, are those that couple naturally to the worldline of
a point particle or the worldvolume of a 6-brane, we see that the
Gibbons-Perry ($N=8,\, D=4$) electric/magnetic duality conjecture may be
re-expressed as an (Type II, $D=10$) particle/sixbrane duality conjecture.
Indeed, the Horowitz-Strominger $D=10$ black sixbrane \host\ is simply obtained
by adding 6 flat dimensions to the $D=4,\, \a=\sqrt{3}$ magnetic black hole.

In general, the $N=8$ theory will admit black holes with $\a=0,1$ and
$\sqrt{3}$ whose extreme limits preserve $1,2$ or $4$ spacetime
supersymmetries,
respectively. Defining ${\cal M}_1=M,\ g_1^2=4\pi Q^2$ and $\kappa^2
=8\pi G$, these extreme black holes satisfy the ``no-force'' condition,
i.e. they saturate the Bogomol'nyi bounds
\eqn\bogo{G(M^2+\Sigma^2)=(1+\a^2)GM^2=N'GM^2=Q^2}
where $\Sigma=\a M$ is the scalar charge and $N'$ is the number of unbroken
supersymmetries.

\newsec{$\a=\sqrt{3}$ in the Heterotic String: Electric/Magnetic Duality in
$D=4$ from String/Fivebrane Duality in $D=10$}

The results of the present paper now allow us to discuss the $\a=\sqrt{3}$
electric/magnetic duality from a totally different perspective from that in
\div\ and section 4.
For concreteness, let us focus on  generic toroidal compactification of the
heterotic string. Instead of the $N=8$ supergravity
of section 4, the four-dimensional theory is now
$N=4$ supergravity coupled to 22  $N=4$ vector
multiplets\footnote{$^\dagger$}
{Gibbons discusses both the $\a=1$ black hole of pure $N=4$
supergravity and the $\a=\sqrt{3}$ Kaluza-Klein black hole in the same paper
\gib, as does Horowitz \hor.
Moreover, black
holes in pure $N=4$ supergravity are treated by Kallosh {\it et al.}
\refs{\klopv\kal{--}\ko}.
The reader may therefore wonder why the $\a=\sqrt{3}$ $N=4$ black hole
discussed in the present paper was overlooked. The reason is that pure $N=4$
supergravity does not admit the $\a=\sqrt{3}$ solution; it is crucial that
we include the $N=4$ vector multiplets in order to introduce the modulus
fields.}.
The same dual Lagrangians \redact\ and \duact\ still emerge but with
completely different origins. The Maxwell field $F_{\mu\nu}$ (or $\tilde{F}_
{\mu\nu}$) and the scalar field $\phi$ do not come from the $D=10$ $2$-form (or
$8$-form) and dilaton of the Type II particle (or sixbrane), but rather from
the $D=10$ $3$-form (or $7$-form) and dilaton plus modulus field of the
heterotic string (or heterotic fivebrane). Thus, the $D=4$ electric/magnetic
duality can now be re-interpreted as a $D=10$ string/fivebrane
duality!

Because of the non-vanishing modulus field  $g_{44}=e^{-2\sigma}$ however,
the $D=10$ black fivebrane solution is not obtained by adding 6 flat
dimensions to the $D=4$ black hole. Rather the two are connected by wrapping
the fivebrane around 5 of the 6 extra dimensions \dufkhu.

The possibility of a heterotic stringy explanation of four-dimensional
electric/magnetic duality has also been considered by Sen
\refs{\sennu\sentwo{--}\senph}. But
his notion of duality is different from the one considered here. The
compactified heterotic string displays a target space duality $O(6,22,Z)$.
It is also conjectured to display the strong/weak coupling
$SL(2,Z)$ $S$-duality relating the dilaton and the axion, which is certainly
there in the field theory limit. See \schwarz\ for a recent review.
The ``duality of dualities'' suggestion
\refs{\schw\bin{--}\schen,\dufkhu} is
that, under string/fivebrane duality, the roles of $S$ and $T$ dualities are
interchanged. Thus Sen's duality refers to $SL(2,Z)$ $S$ duality, whereas
ours refers to $S\leftrightarrow T$. A similar distinction is to be drawn
between the two uses of the words ``dual string'' \refs{\sentwo,\dufkhu}.
In any event, the picture that emerges is one in which the massive states
of the string correspond to extreme black holes.

\newsec{Conclusion}
Of course, the present paper has established only that these two-parameter
configurations are solutions of the field theory limit of the heterotic string.
Although the extreme one-parameter solutions are expected to be exact to all
orders in $\alpha'$, the same reasoning does not carry over to the new
two-parameter solutions. It would be interesting to pursue conformal field
theory arguments, perhaps along the lines recently suggested by Giddings,
Polchinski and Strominger \gps.

It would be also interesting to see whether the generalization of the
one-parameter solutions of \dufkhu\  to the two-parameter solutions of the
present paper can be carried out when we include the Yang-Mills coupling as in
\dufkhu. This would necessarily involve giving up
the self-duality condition on the Yang-Mills field strength, however, since the
self-duality condition is tied to the extreme, $\sqrt{2} \kappa {\cal M}_{p+1}
=  g_{p+1}$, supersymmetric solutions.

Finally, there is the question of whether these solutions are peculiar
to the toroidal compactification or whether they survive in more realistic
orbifold or Calabi-Yau models \dauria. Although the actions $S_{1}$, $S_{2}$
and $S_{3}$ were originally derived in the context of the torus \dufkhu, they
also appear in a large class of $N=1$ supergravity theories.

\vskip20pt
{\bf Acknowledgments}

We have enjoyed useful conversations with Ergin Sezgin.
\vfil\eject
\listrefs
\bye